\documentclass[a4paper,11pt]{article}
\usepackage{amssymb}
\usepackage{amsmath}
\usepackage{amsfonts}
\usepackage{mathtools}
\usepackage[noend]{algpseudocode,algorithm}
\usepackage{slashed}
\usepackage{color}
\usepackage{indentfirst}
\usepackage{graphicx}
\usepackage{bbm}
\usepackage{csquotes} 
\usepackage{caption}
\usepackage{cite}
\usepackage{here}
\usepackage{comment}
\usepackage{listings}
\numberwithin{equation}{section} 

\makeatletter
\def\BState{\State\hskip-\ALG@thistlm}
\makeatother
\makeatletter
\renewcommand{\ALG@name}{Algorithm}
\makeatother
\algrenewcommand\algorithmicdo{}

\renewcommand{\thefootnote}{\fnsymbol{footnote}}

\begin{document}

\begin{titlepage}

\hfill{KEK-TH-2306}

\begin{center}
\vspace*{1cm}
{\Large \bf
Small flux superpotential in F-theory compactifications}
\vskip 1.5cm
{\large Yoshinori Honma${}^{a}$\footnote[2]{yhonma@law.meijigakuin.ac.jp} and 
Hajime Otsuka${}^b$\footnote[3]{hotsuka@post.kek.jp}}
\vskip 1.0cm
{\it 
${}^a$%
Department of Current Legal Studies, \\
Meiji Gakuin University, \\
Yokohama, Kanagawa 244-8539, Japan \\
\vspace*{0.5cm}
${}^b$%
Theory Center, Institute of Particle and Nuclear Studies, \\
High Energy Accelerator Research Organization, KEK, \\
1-1 Oho, Tsukuba, Ibaraki 305-0801, Japan\\}
\end{center}
\vskip2.5cm


\begin{abstract}
We investigate whether a class of models describing F-theory compactifications admits 
a specific type of flux vacua with an exponentially small vacuum expectation value of the 
superpotential, by generalizing a method recently developed in Type IIB flux 
compactifications. First we clarify that a restricted choice of $G_4$-flux components 
reduces a general flux superpotential into a simple form, which promotes the existence of 
supersymmetric vacua with one flat direction at the perturbative level. Then we utilize the 
techniques of mirror symmetry to determine one-instanton corrections to the potential and 
investigate in detail the vacuum solutions of a particular model.
\end{abstract}


\end{titlepage}
\renewcommand{\thefootnote}{\arabic{footnote}}
\setcounter{footnote}{0}
\clearpage
\addtocounter{page}{0}

\section{Introduction}

Stabilization of moduli fields arising from space-time compactifications is 
of particular relevance for the construction of four-dimensional realistic 
universe from string theory as well as higher-dimensional gravity. Moduli fields 
correspond to not only the size and shape of the extra-dimension, but also determine 
various couplings in the resultant four-dimensional effective theory. In order 
not to be contradicted with the observational constraints obtained from the realm of
cosmology and phenomenology, one needs to fix the dynamics of moduli fields adequately. 

One of the well-established frameworks of moduli stabilization has been 
realized in Type IIB string theory on Calabi-Yau orientifolds. There the Ramond-Ramond and 
Neveu-Schwarz fluxes in string theory stabilize the complex structure moduli and the 
axio-dilaton, while the K$\ddot{{\textrm{a}}}$hler moduli can be stabilized by 
non-perturbative corrections. More precisely, it is quite well known that the 
imaginary self-dual three-form fluxes in Type IIB string theory can provoke stable 
Minkowski minima, on which complex structure moduli and axio-dilaton are stabilized at 
a compactification scale \cite{Giddings:2001yu}. In the Kachru-Kallosh-Linde-Trivedi 
(KKLT) model \cite{Kachru:2003aw} and the LARGE Volume Scenario developed 
in \cite{Balasubramanian:2005zx}, it has been shown that perturbative stringy 
$\alpha^\prime$ corrections and non-perturbative instanton corrections with appropriate 
uplifting mechanism completely fix the remaining K$\ddot{{\textrm{a}}}$hler moduli, 
providing the outline of the construction of de Sitter space-time in string theory 
context. There the smallness of the value of flux-induced superpotential for complex 
structure moduli fields played a crucial role.

Recently, the authors of \cite{Demirtas:2019sip} (see 
also \cite{Demirtas:2020ffz,Blumenhagen:2020ire}) described a feasible method for 
constructing flux vacua with exponentially suppressed superpotential in the framework 
of Type IIB Calabi-Yau orientifolds, which can be naturally incorporated into the 
KKLT construction. The main idea to obtain such a small flux superpotential is to 
utilize non-perturbative terms in the prepotential to the fullest extent. More 
precisely, they first neglected non-perturbative corrections in the prepotential and 
made a restricted choice of background fluxes such that the possible minima at the 
perturbative level admit a flat direction along with exactly vanishing superpotential. 
Then, by taking into account the non-perturbative corrections to the effective theory, 
the remaining modulus acquires a small mass and they found preferable minima with 
exponentially small flux superpotential. This mechanism is so simple and thus have a 
broad range of applicability, especially toward explicit constructions of KKLT-like 
scenario in various effective field theories.

In this paper, we evaluate the validity of the above method in the framework of
F-theory flux compactifications, where both the open and closed string moduli fields 
can be geometrically controlled \cite{Vafa:1996xn}. Extending the Type IIB setup 
of \cite{Demirtas:2019sip} to F-theory compactified on Calabi-Yau fourfolds, we clarify 
how to obtain a specific type of perturbatively flat vacua with exactly vanishing 
superpotential. After specifying an explicit example for the background, we utilize the 
mirror symmetry of Calabi-Yau fourfolds to compute relevant non-perturbative 
corrections to the potential. To solve intricate equations of the system appropriately, 
we rely on a numerical calculation and investigate whether a desirable class of 
F-theory vacua with small superpotential can be allowed in our setup. Our verification 
may shed new light on a realization of KKLT-like scenario over a broad range of stringy 
frameworks. 

This paper is organized as follows. First we briefly look at a general formula about the
effective theory arising from F-theory compactifications in Section \ref{sec:2}. 
Then we pick up a particular Calabi-Yau fourfold as an explicit example, and numerically 
investigate the model to verify the existence of flux vacua equipped with the
exponentially small superpotential in Section \ref{sec:3}. Section \ref{sec:con} is 
devoted to conclusions and discussions. In the Appendix A, we describe a detailed discussion 
about various perturbative vacuum solutions of our example.

\section{F-theory compactifications}
\label{sec:2}

Here we describe several basic ingredients about space-time compactifications in F-theory 
framework. For more details, we refer the reader to \cite{Denef:2008wq} and references 
therein. We also outline our strategy to find out a class of perturbatively flat vacua, 
which becomes a key ingredient to realize the small vacuum expectation value of the flux 
superpotential in a subsequent discussion.

\subsection{Basic setup}
\label{subsec:Fsetup}

Here we first take a brief look at general aspects of four-dimensional ${\cal{N}}=1$ 
effective theory arising from F-theory compactified on a Calabi-Yau fourfold $X_4$. In 
terms of the four-dimensional ${\cal{N}}=1$ supersymmetry, effective interactions of moduli
fields can be determined by the K$\ddot{{\textrm{a}}}$hler potential and superpotential.
Calabi-Yau compactifications enable us to derive these ingredients quantum mechanically 
from characteristics of the underlying Calabi-Yau manifolds. 

More precisely, the K$\ddot{{\textrm{a}}}$hler potential for complex structure moduli 
space of $X_4$ is defined by
\begin{align}
K=-\ln{\int_{X_4}} \Omega \wedge \overline{\Omega},
\label{Kahp}
\end{align}
where $\Omega$ denotes a holomorphic $(4,0)$-form on $X_4$. Here and in what follows, we 
adopt the reduced Planck unit $M_{\rm Pl}=2.4\times 10^{18}\,{\rm GeV}=1$.  In the
presence of background four-form fluxes $G_4$, Calabi-Yau compactifications also admit a 
flux superpotential of the form \cite{Gukov:1999ya}
\begin{align}
W = \int_{X_4} G_4 \wedge \Omega,
\end{align} 
which is inherited from a duality between the F-theory and M-theory compactified on 
the same manifold \cite{Becker:1996gj,Sethi:1996es,Dasgupta:1999ss,Haack:2001jz}.
Here background fluxes are required to satisfy the tadpole cancellation condition 
given by 
\begin{align}
\frac{\chi}{24}=n_{\rm{D3}}+\frac{1}{2}\int_{X_4} G_4 \wedge G_4,
\label{tad}
\end{align}
in order to globally conserve the total charges within a compact manifold $X_4$. Here 
$\chi$ represents the Euler characteristic of $X_4$ and $n_{\rm{D3}}$ denotes the total 
number of mobile D3-branes freely moving in $X_4$. 

For a Calabi-Yau fourfold $X_4$ equipped with $h^{3,1}(X_4)$ complex structure moduli, the 
so-called period integrals of holomorphic $(4,0)$-form defined by
\begin{align}
\Pi_A = \int_{\gamma_A} \Omega,
\label{4peri}
\end{align}
encode the moduli dependence of the effective theory, and in particular the 
K$\ddot{{\textrm{a}}}$hler potential (\ref{Kahp}) for the complex structure moduli 
fields can be re-expressed as
\begin{align}
e^{-K} = \Pi \cdot \eta \cdot \overline{\Pi},
\label{Kper}
\end{align}
where $\gamma_A$ with $A=1, \ldots , h^4_H (X_4)$ correspond to basis elements of primary 
horizontal subspace of $H_4 (X_4)$. An intersection matrix $\eta_{AB}$ and a dual 
basis $\hat{\gamma}^A$ in $H^4_H (X_4)$ are defined accordingly by
\begin{align}
\eta^{AB} = \int_{X_4} \hat{\gamma}^A \wedge \hat{\gamma}^B,
 \ \ \ \ \ \ \ \int_{\gamma_A} \hat{\gamma}^B = \delta_A^{ \ B}. 
\label{intm} 
\end{align}
Similarly, if the underlying internal manifolds are filled with background four-form 
fluxes whose integer quantum numbers are given by
\begin{align}
n_A = \int_{\gamma_A} G_4,
\label{Gflux}
\end{align}
moduli fields have interactions through the following form of the flux superpotential 
\begin{align}
W = n_A \ \eta^{AB} \ \Pi_B,
\label{fluq}
\end{align}
in terms of period integrals.

\subsection{Our strategy}
\label{subsec:Fstrat}

Throughout this paper, we focus on a particular class of F-theory flux compactifications 
whose underlying Calabi-Yau fourfolds $X_4$ around a large complex structure point provide 
a specific type of moduli dependence into $W$ given by 
\begin{align}
\begin{split}
W &= C_0(n_A)+\widetilde{C}_0(n_A) S+C_a(n_A) z^a+\widetilde{C}_a(n_A) S z^a
+\frac{1}{2}C_{ab}(n_A) z^az^b+\frac{1}{2}\widetilde{C}_{ab}(n_A) Sz^az^b \\
& \ \ \ +\frac{1}{3!}C_{abc}(n_A) z^az^bz^c+\frac{1}{3!}\widetilde{C}_{abc}(n_A) Sz^az^bz^c
+\frac{1}{4!}C_{abcd}(n_A) z^az^bz^cz^d,
\end{split}
\label{fsupans}
\end{align}
at the classical level.\footnote{In other words, here we assume that the background manifold 
has a certain fibration structure whose fiber has intersection number 0 with itself.} 
Here $a = 1, 2, \cdots, h^{3,1}(X_4)-1$ and the coefficients 
\begin{align}
\{C_0, \widetilde{C}_0, C_a, \widetilde{C}_a, C_{ab}, \widetilde{C}_{ab}, C_{abc}, 
\widetilde{C}_{abc}, C_{abcd}\}
\end{align}
are functions of the background $G_4$-fluxes $n_A$, whose explicit form can be determined 
from topological data of $X_4$. Most important property of the above ansatz for 
F-theory superpotential is the existence of a particular moduli field $S$ which can enter 
only linearly in each terms of the polynomial. As we will show in the next section, we 
realize $S$ by a field originating from either the axio-dilaton or a linear combination 
of axio-dilaton and other moduli in the language of Type IIB string theory.

Adopting the prescription developed 
in \cite{Demirtas:2019sip,Demirtas:2020ffz,Blumenhagen:2020ire} into our present setup, 
it turns out that if there exists a restricted choice of $G_4$-fluxes whose non-zero 
contributions include primitive (2,2)-components (see for instance \cite{Denef:2008wq}) 
on a certain background such that the superpotential becomes a simple form\footnote{In our 
approach, we started from just picking up homogeneous of degree two terms of the 
superpotential to realize the small $W$, which also simultaneously results in a discrete 
shift symmetric formula. Our prescription corresponds to provide a sufficient condition 
in the context of the original discussion in Type IIB setup \cite{Demirtas:2019sip,Demirtas:2020ffz},
and it would be interesting to clarify a necessary condition for the derivation along the lines
of \cite{Demirtas:2019sip,Demirtas:2020ffz}.}
\begin{align}
W=\widetilde{C}_a S z^a +\frac{1}{2}C_{ab} z^az^b,
\label{desf}
\end{align}
a class of supersymmetric F-theory vacua satisfying
\begin{align}
\partial_S W = \partial_a W = W = 0,
\label{vsol}
\end{align}
can perturbatively exist along a one-dimensional locus
\begin{align}
z^a = S P^a,
\label{Floc}
\end{align}
if 
$P^a \equiv -\frac{1}{2} (C^{-1})^{ab} \ \widetilde{C}_b$ satisfy the following 
condition:
\begin{align}
\widetilde{C}_a P^a=0.
\label{eq:conditionF}
\end{align}

Although stringy $\alpha^\prime$ corrections generically induce extra contributions 
proportional to $\zeta (3)$ into (\ref{fsupans}) at the perturbative 
level \cite{Honma:2013hma}, the restricted choice of four-form fluxes picking up homogeneous 
of degree two terms does not support such a deformation, 
leading to the same result (\ref{desf}). 

From a geometric point of view, the reduced form (\ref{desf}) can be realized from a certain 
combination of 3-point topological Yukawa couplings of the underlying Calabi-Yau fourfold 
explicitly given by
\begin{align}
\frac{1}{2}n_{\alpha}\eta^{\alpha\beta} z^iz^j \int_{X_4} H_{\beta} \wedge J_i \wedge J_j,
\end{align}
where $\alpha$ labels the elements of primary subspace of the cohomology 
$H^{2,2}_{\textrm{prim}}(X_4) \subset H^{2,2}(X_4)$ whose bases $H_{\beta}$ are generated 
by the wedge products of the K$\ddot{{\textrm{a}}}$hler form $J_i \wedge J_j$. 
Correspondingly $n_{\alpha}$ represent (2,2)-components 
of the background $G_4$-flux quanta. Here we have assumed the existence of the appropriate 
modulus $S$ and utilized a unified expression for moduli space parameters as $z^i \equiv
\{z^a,S\}$.

Once there exists an appropriate background admitting perturbative solutions of the 
above type, there would be a possibility for constructing desirable vacua where only 
the instanton corrections can contribute to the mass of the remaining modulus and 
induce a small superpotential, realizing a stable minima equipped with preferable
properties.\footnote{Since the identification of the modulus $S$ and determination of 
its appearance in flux superpotential strongly depends on the choice of manifolds and 
its triangulation, seeking a classification for suitable or unsuitable geometries 
for the mechanism would be quite difficult. However, there is no doubt that establishing 
a certain classification problem may provide a wide perspective about the subject.}

\section{An explicit example}
\label{sec:3}

In the remaining part of this paper, we focus on a specific elliptically-fibered Calabi-Yau
fourfold to exemplify the realization of exponentially small vacuum expectation value of 
F-theory flux superpotential.

\subsection{Perturbatively flat vacua}
\label{subsec:P11169set}

First we will show a detailed setup of our demonstration and elucidate the existence of the 
perturbatively flat solution of the above type, which becomes a key ingredient to realize 
desirable vacua after including non-perturbative corrections to the system.

Let us consider a mirror pair ($X_4, \tilde{X}_4$) of Calabi-Yau fourfolds studied 
in \cite{Grimm:2010gk},\footnote{Here we would like to comment that we have also evaluated 
another simple Calabi-Yau fourfold studied in \cite{Honma:2017uzn,Honma:2019gzp} and 
confirmed that there are no desirable perturbative solutions with flat directions, 
largely because in that case the tadpole cancellation condition becomes too severe.} whose 
associated toric charge vectors \cite{Batyrev:1993dm,Batyrev:1994pg} are given by\footnote{See
\cite{Alim:2009bx,Grimm:2009ef,Jockers:2009ti,Grimm:2010gk,Honma:2015iza} for the details 
about explicit constructions of the background based on mirror symmetry techniques.}
\begin{align}
\begin{split}
    \ell^1 &= ( \ \ 0, -2, \ \ 1, \ \ 0, \ \ 1, \ \ 0, \ \ 0, \ \ 1, -1, \ \ 0), \\
    \ell^2 &= (-6, \ \ 1, \ \ 0, \ \ 0, \ \ 0, \ \ 2, \ \ 3, \ \ 0, \ \ 0, \ \ 0), \\
    \ell^3 &= ( \ \ 0, -1, \ \ 0, \ \ 1, \ \ 0, \ \ 0, \ \ 0, -1, \ \ 1, \ \ 0), \\
    \ell^4 &= ( \ \ 0, -1, \ \ 0, -1, \ \ 0, \ \ 0, \ \ 0, \ \ 1, \ \ 0, \ \ 1).
\label{torvec4}
\end{split}
\end{align}
Their topological quantities such as the Hodge numbers and Euler characteristic are 
\begin{align}
\begin{split}
    &h^{3,1}(X_4)=h^{1,1}(\tilde{X}_4) =4, \\
    &h^{2,1}(X_4)=h^{2,1}(\tilde{X}_4) =0, \\
    &h^{1,1}(X_4) =h^{3,1}(\tilde{X}_4) =2796, \\
    &h^{2,2}(X_4)=h^{2,2}(\tilde{X}_4) =11244, \\
    &\chi (X_4)=\chi(\tilde{X}_4) = 16848,
\end{split}
\end{align}
and the dimension of primary horizontal subspace of $H_4 (X_4)$ is given by 
$h^4_H (X_4) = 16$, which also determines the total number of elements of the 
independent background fluxes $n_A$.

After a standard mirror symmetry calculation with the charge assignment in 
(\ref{torvec4}), one can show that there exist 16 independent period integrals 
associated with $X_4$ whose perturbative expansions are given by\footnote{Note that 
our present notation for the complex structure moduli fields $\{z \}$ deviates from 
a standard convention where the classical periods are expressed by logarithmic 
functions of the complex structure deformations. We have redefined such a logarithm 
of a standard complex structure modulus as a new single modulus, just for the later 
convenience.}
\begin{align}
\begin{split}
\Pi_1 &= 1,
\quad \Pi_2 = z_1,
\quad \Pi_3 = z_2,
\quad \Pi_4 = z_3,
\quad \Pi_5 = z_4, \\
\Pi_6 &= z_2 (z_2+z_3+z_4),
\quad \Pi_7 = z_2 (2z_1+3z_2+2z_3),
\quad \Pi_8 = 2z_2 (z_1+2z_2+z_3+z_4), \\
\Pi_9 &=  2z_1^2 +24z_1z_2 +\frac{103}{2}z_2^2 +9z_1z_3 +34z_2z_3 
+\frac{9}{2}z_3^2 +7(z_1+3z_2+z_3)z_4, \\
\Pi_{10} &=  \frac{1}{2}\biggl[ 3z_1^2 +28z_1z_2 +55z_2^2 +10z_1z_3 
+36z_2z_3 +5z_3^2 +6(z_1+3z_2+z_3)z_4\biggl], \\
\Pi_{11} &=  \frac{1}{2}\biggl[ 3z_1^2 +30z_1z_2 +61z_2^2 +10z_1z_3 
+38z_2z_3 +5z_3^2 +8(z_1+3z_2+z_3)z_4\biggl], \\
\Pi_{12} &=  -\frac{z_2}{6}\biggl[ 11z_2^2 +6z_1(z_2 +z_3 +z_4) 
+3z_3(z_3 +2z_4) +3z_2(4z_3+3z_4)\biggl] 
+\frac{165i \zeta(3)}{2\pi^3}, \\
\Pi_{13} &=  -\frac{z_2}{6}\left( 6z_1^2 +33z_1z_2 +46z_2^2\right) 
+\frac{1}{2}(z_1 +3z_2)(z_1+5z_2)z_3 
+\frac{1}{2}(z_1 +4z_2)z_3^2+\frac{z_3^3}{6} \\
&  \ \ \ \ +\frac{z_4}{2}(z_1 +3z_2+z_3)^2+\frac{347i \zeta(3)}{\pi^3}, \\
\Pi_{14} &=  -\frac{z_2}{2}\biggl[ z_1^2 +5z_2^2 +4z_2z_3 +z_3^2 +3z_2z_4 
+2z_3z_4 +2z_1(2z_2 +z_3 +z_4)\biggl] 
+\frac{225i \zeta(3)}{2\pi^3}, \\
\Pi_{15} &=  -\frac{z_2}{2}\left( z_1^2 +3z_1z_2+3z_2^2 +2z_1z_3 +3z_2z_3 
+z_3^2\right) +\frac{135i \zeta(3)}{2\pi^3}, \\
\Pi_{16} &=  \frac{z_2}{12}\biggl[ 22z_1z_2^2 +23z_2^3 +6z_1^2(z_2 +z_3 +z_4) 
+6z_1z_3(z_3 +2z_4)+2z_3^2(z_3 +3z_4)+6z_2z_3(2z_3 +3z_4) \\
& \ \ \ \ +6z_1z_2(4z_3 +3z_4)+6z_2^2(5z_3 +3z_4)\biggl] 
-\frac{i (165z_1 +694z_2 +225z_3 +135z_4) \zeta(3)}{2\pi^3},
\label{Picl}
\end{split}
\end{align}
around a large complex structure point $z_{1,2,3,4} = \infty$. Here we 
have abbreviated further possible corrections originating from worldsheet 
instantons in the topological A-model, in order to restrict ourselves, at this
stage, to the perturbative analysis.

In this fourfold example, the associated intersection matrix defined in (\ref{intm}) 
becomes
\begin{align}
\eta=\begin{pmatrix}
0& 0 & 0 & 0 & 1 \\
0 & 0 & 0 & I_4 & 0 \\
0 & 0 & \widetilde{\eta} & 0 & 0 \\
0 & I_4 & 0 & 0 & 0 \\
1 & 0 & 0 & 0 & 0
\end{pmatrix},
\label{imat}
\end{align}
\textrm{with}
\begin{align}
\widetilde{\eta}=\begin{pmatrix}
2& 0 & -1 & -\frac{5}{7} & \frac{1}{7} & \frac{8}{7} \\[5pt]
0& \frac{1}{2} & -\frac{3}{4} & 0 & -\frac{1}{2} & \frac{1}{2} \\[5pt]
-1& -\frac{3}{4} & \frac{1}{2} & \frac{3}{14} & \frac{5}{14} & -\frac{9}{14} \\[5pt]
-\frac{5}{7}& 0 & \frac{3}{14} & 0 & 0 & 0 \\[5pt]
\frac{1}{7} & -\frac{1}{2} & \frac{5}{14} & 0 & 0 & 0 \\[5pt]
\frac{8}{7} & \frac{1}{2} & -\frac{9}{14} & 0 & 0 & 0
\end{pmatrix},
\end{align}
and the explicit form of the flux superpotential is given by
\begin{align}
\begin{split}
W &= n_{16}+n_{15}z_4+n_{12}z_1+n_{13}z_2+n_{14}z_3+\frac{z_4}{14}
(7n_7z_1+12n_{10}z_2-2n_{11}z_2-4n_9z_2+7n_7z_3) \\
& \quad +\frac{1}{28}(14n_6 z_1^2-8n_{10}z_1z_2-8 n_{11}z_1z_2+12 n_9 z_1z_2
+2 n_{10}z_2^2+2n_{11}z_2^2+4n_9z_2^2+14n_8z_1z_3 \\
& \quad -4n_{10}z_2z_3+24n_{11}z_2z_3-8n_9z_2z_3+7n_8z_3^2)
-\frac{z_4}{2}(n_3z_1^2+2n_2z_1z_2+6n_3z_1z_2+2n_4z_1z_2 \\
& \quad +3n_2z_2^2+9n_3z_2^2+3n_4z_2^2+2n_3z_1z_3+2n_2z_2z_3
+6n_3z_2z_3+2n_4z_2z_3+n_3z_3^2) \\
& \quad -\frac{1}{6}(3n_4 z_1^2z_2+6n_3 z_1^2 z_2+3n_5 z_1^2z_2+6n_2 z_1z_2^2
+33n_3z_1z_2^2+12n_4z_1z_2^2+9n_5z_1z_2^2+11n_2z_2^3 \\
& \quad +46n_3z_2^3+15n_4z_2^3+9n_5z_2^3+3n_3z_1^2 z_3+6n_2z_1z_2z_3
+24n_3z_1z_2z_3+6n_4z_1z_2z_3+6n_5z_1z_2z_3 \\
& \quad +12n_2z_2^2 z_3+45n_3z_2^2 z_3+12n_4z_2^2z_3+9n_5z_2^2z_3+3n_3z_1z_3^2
+3n_2z_2z_3^2+12n_3z_2z_3^2+3n_4z_2z_3^2 \\
& \quad +3n_5z_2z_3^2+n_3z_3^3)+\frac{z_4}{2}(n_1z_1^2 z_2+3n_1z_1z_2^2
+3n_1z_2^3+2n_1z_1z_2z_3+3n_1z_2^2 z_3+n_1z_2z_3^2) \\
& \quad +\frac{1}{12}(6n_1z_1^2 z_2^2+22n_1z_1z_2^3+23n_1z_2^4+6n_1z_1^2 z_2z_3
+24n_1z_1z_2^2z_3+30n_1z_2^3z_3+6n_1z_1z_2z_3^2\\
& \quad +12n_1z_2^2z_3^2+2n_1z_2z_3^3)+\frac{i\zeta(3)}{2\pi^3}
\biggl[165n_2+694n_3+225n_4+135n_5-n_1(165z_1+694z_2 \\
& \quad +225z_3+135z_4)\biggl],
\end{split}
\label{exfs}
\end{align}
which enables us to identify the modulus $z_4$ inherited from the axio-dilaton as $S$ 
explained in Section \ref{subsec:Fstrat}. Note that we rewrite $z_4$ as $S$ in 
subsequent discussions. At the perturbative level, the dynamics of 
effective theory arising from F-theory compactified on $X_4$ can be completely 
specified by (\ref{exfs}) and the K$\ddot{{\textrm{a}}}$hler potential (\ref{Kper}) 
explicitly given by
\begin{align}
\begin{split}
    K &= (S - \bar{S})
\biggl[
  \frac{(z_1 - \bar{z}_1)^2 (z_2 - \bar{z}_2)}{2} + 
 \frac{3(z_1 - \bar{z}_1) (z_2 - \bar{z}_2)^2}{2} + 
 \frac{3 (z_2 - \bar{z}_2)^3}{2}
 + (z_1 - \bar{z}_1) (z_2 - \bar{z}_2) (z_3 - \bar{z}_3) \\
 & \quad +\frac{3(z_2 - \bar{z}_2)^2 (z_3 - \bar{z}_3)}{2} + 
 \frac{(z_2 - \bar{z}_2) (z_3 - \bar{z}_3)^2}{2}\biggl]
+\frac{(z_1 - \bar{z}_1)^2 (z_2 - \bar{z}_2)^2}{2} + \frac{11 (z_1 - \bar{z}_1)
(z_2 - \bar{z}_2)^3}{6} \\
 & \quad +\frac{23 (z_2 - \bar{z}_2)^4}{12}+\frac{(z_1 - \bar{z}_1)^2 (z_2 - \bar{z}_2)
 (z_3 - \bar{z}_3)}{2}+2 (z_1 - \bar{z}_1) (z_2 - \bar{z}_2)^2 (z_3 - \bar{z}_3) \\
 & \quad +\frac{5 (z_2 - \bar{z}_2)^3 (z_3 - \bar{z}_3)}{2} + 
  \frac{(z_1 - \bar{z}_1) (z_2 - \bar{z}_2) (z_3 - \bar{z}_3)^2}{2}+(z_2 - \bar{z}_2)^2
  (z_3 - \bar{z}_3)^2 \\
  & \quad +\frac{(z_2 - \bar{z}_2) (z_3 - \bar{z}_3)^3}{6}
  +\frac{i \zeta(3)}{4 \pi^3}\left[ -660 (z_1 - \bar{z}_1)-540(S-\bar{S})
  -900(z_3-\bar{z}_3)-2776(z_2-\bar{z}_2) \right].
\end{split}
\label{expliK}
\end{align}

Now let us check whether the above example of F-theory flux compactifications 
appropriately admits a class of perturbative supersymmetric solution with a flat 
direction discussed in the previous section. When the background four-form $G_4$-fluxes 
are restricted to be of the type (2,2) and primitive as represented by setting
\begin{align}
    n_1 = n_2 = n_3 = n_4 = n_5 = n_{12} = n_{13} = n_{14} = n_{15} = n_{16} = 0,
\end{align}
one can easily show that the flux superpotential of the model becomes a desirable 
form (\ref{desf}) with the following assignments:
\begin{align}
\widetilde{C}_1 =\frac{n_7}{2}, \quad \widetilde{C}_2=\frac{1}{7}(6n_{10}-n_{11}-2n_9), 
\quad \widetilde{C}_3=\frac{n_7}{2},
\end{align}
and
\begin{align}
\begin{pmatrix}
C_{11} & C_{12} & C_{13}  \\
C_{21} & C_{22} & C_{23}  \\
C_{31} & C_{32} & C_{33}
\end{pmatrix}
=\begin{pmatrix}
n_6 & \frac{-2n_{10} -2 n_{11} +3 n_9}{7} & \frac{n_8}{2}  \\
\frac{-2n_{10} -2 n_{11} +3 n_9}{7} & \frac{n_{10} + n_{11}+ 2 n_9}{7} & -\frac{n_{10} 
- 6 n_{11} + 2 n_9}{7}  \\
\frac{n_8}{2} & -\frac{n_{10} - 6 n_{11} + 2 n_9}{7} & \frac{n_8}{2}
\end{pmatrix}.
\label{cabm}
\end{align}

Note that the non-zero (2,2)-components of the background fluxes must be chosen such that 
the matrix $C_{ab}$ in (\ref{cabm}) continues to be invertible and the condition 
(\ref{eq:conditionF}) with the assignments
\begin{align}
\footnotesize
\begin{split}
 P^1&=\frac{1}{{\cal A}}(n_{10} + 8 n_{11} - 5 n_9) (-n_{11} (6 n_7 + n_8) 
 + n_{10} (n_7 + 6 n_8) + 2 (n_7 - n_8) n_9), \\
P^2&=\frac{7}{2{\cal A}} (2 n_6 - n_8) (-n_{11} (6 n_7 + n_8) + n_{10}(n_7 + 6 n_8) 
+ 2 (n_7 - n_8) n_9), \\
P^3&=\frac{2 n_{10}^2}{{\cal A}} (6 n_6 - n_7 - 6 n_8) + \frac{2 n_{11}^2}{{\cal A}} 
(6 n_6 - 8 n_7 + n_8) + 
   \frac{n_{11}}{2{\cal A}} (14 n_6 n_7 - 7 n_7 n_8 + 40 n_6 n_9 + 68 n_7 n_9 + 2 n_8 n_9)
   \\
   &- \frac{n_9}{{\cal A}} (-14 n_6 n_7 + 7 n_7 n_8 + 8 n_6 n_9 + 15 n_7 n_9 + 6 n_8 n_9)
   \\
   &+ \frac{n_{10}}{2{\cal A}} (14 n_6 n_7 - 7 n_7 n_8 - 4 n_{11} (37 n_6 + 9 n_7 + 5 n_8) 
   + 40 n_6 n_9 +26 n_7 n_9 + 44 n_8 n_9), \\
{\cal A} &\equiv 
4 n_{10}^2 n_6 + 8 n_{11}^2 (18 n_6 + 7 n_8) + 7 n_{11} n_8 (n_8 - 16 n_9) - 
 2 n_{11} n_6 (7 n_8 + 48 n_9) + 
 2 n_9 (7 n_8 (n_8 + 3 n_9) + n_6 (-14 n_8 + 8 n_9))\\
 &+ 
 n_{10} (8 n_{11} (-6 n_6 + 7 n_8) + 7 n_8 (n_8 - 4 n_9) + 2 n_6 (-7 n_8 + 8 n_9)),
\label{eq:pi}
\end{split}
\end{align}
\normalsize
are also satisfied.\footnote{Here we have omitted the possibility of another 
type of perturbatively flat solution satisfying $\textrm{Det} (C_{ab})=0$ and 
more detailed analysis of the vacuum solutions of the present example are relegated to 
the Appendix \ref{subsec:Noninv}.}

Here let us pick up a particular choice of background fluxes whose non-zero components 
are 
\begin{align}
    n_6 = -10,\quad n_7=n_{11}=-8,\quad
    n_8= 12, \quad n_9 = 7,\quad n_{10}=15.
\end{align}
In order to satisfy the tadpole cancellation condition (\ref{tad}) of the present 
model, the background also needs to contain additional $n_{D3}=2$ mobile D3-branes, 
whose numbers are sufficiently small to guarantee our exclusion of the backreaction 
to the space-time. After plugging these inputs into the general formula (\ref{Floc}), 
one finds that the model admits a desirable minimum equipped with a perturbative flat 
direction along
\begin{align}
    \left(
    \begin{array}{c}
         z_1  \\
         z_2 \\
         z_3
    \end{array}
    \right) =
    \frac{S}{7}\left(
    \begin{array}{c}
         3  \\
         4 \\
         9
    \end{array}
    \right).
    \label{eq:flatex}
\end{align}
This solution corresponds to the ``Vacuum A" specified in the next section.

\subsection{Non-perturbative uplifting of flat direction}
\label{subsec:P11669qu}

By using algebraic methods of the toric geometry (see for 
example \cite{Hosono:1993qy,Hosono:1994ax}), one can determine non-perturbative 
quantum corrections in the moduli space of Calabi-Yau manifolds by solving the associated 
Picard-Fuchs equations and calculating the period integrals explicitly. For the fourfold
example described in the previous subsection, it turns out that the corresponding period
integrals (\ref{Picl}) take the following form at the non-perturbative level:
\begin{align}
    \widetilde{\Pi}_1 &= \Pi_1=1,\quad \widetilde{\Pi}_2 =\Pi_2= z_1,
    \quad
    \widetilde{\Pi}_3 =\Pi_3= z_2,
    \quad
    \widetilde{\Pi}_4 =\Pi_4= z_3,
    \quad
    \widetilde{\Pi}_5 =\Pi_5= S,
    \nonumber\\
    \widetilde{\Pi}_6 &= \Pi_6+\frac{e^{2 \pi i z_1}}{4\pi^2},
    \quad
    \widetilde{\Pi}_7 = \Pi_7-\frac{e^{2 \pi i S}}{2\pi^2},
    \quad
    \widetilde{\Pi}_8 = \Pi_8-\frac{e^{2 \pi i z_3}}{2\pi^2},
    \quad
    \widetilde{\Pi}_9 = \Pi_9 
    -\frac{1545e^{2 \pi i z_2}}{\pi^2},
    \nonumber\\
    \widetilde{\Pi}_{10} &= \Pi_{10} 
    -\frac{825e^{2 \pi i z_2}}{\pi^2},
    \quad
    \widetilde{\Pi}_{11} = \Pi_{11} 
    -\frac{915e^{2 \pi i z_2}}{\pi^2},
    \nonumber\\
\widetilde{\Pi}_{12} &= \Pi_{12}+\frac{1}{8\pi^3}
    \biggl[ e^{2\pi i z_1}(i +2\pi z_1) -1320e^{2\pi i z_2}(i +\pi z_2) 
    -2e^{2\pi i z_3} (i +\pi z_3) -2e^{2\pi i S}(i +\pi S)
    \biggl],
    \nonumber\\
\widetilde{\Pi}_{13} &= \Pi_{13}-\frac{15e^{2 \pi i z_2}}{\pi^3}
    \biggl[ 23i +\pi (11 z_1 + 46 z_2 + 15 z_3 + 9 S)
    \biggl],
    \nonumber\\
\widetilde{\Pi}_{14} &= \Pi_{14}-\frac{1}{8\pi^3}
    \biggl[ 1800e^{2\pi i z_2}(i +\pi z_2)+e^{2\pi i z_3}(i + 2 \pi (z_1 + z_3)) 
    +2e^{2\pi i S} (i +\pi S)
    \biggl],
    \nonumber\\
\widetilde{\Pi}_{15} &= \Pi_{15}
   -\frac{1}{4\pi^3}
    \biggl[  540e^{2\pi i z_2} (i +\pi z_2)+e^{2\pi i S}\pi (z_1 + z_3)
    \biggl],
    \nonumber\\
\widetilde{\Pi}_{16} &=  \Pi_{16}-\frac{1}{8\pi^3}
    \biggl[ -e^{2\pi i z_1}z_1(i + \pi z_1)
    +e^{2\pi i z_3}(2z_1 +z_3)(i +\pi z_3)
    +2e^{2\pi i S}(z_1+z_3)(i + \pi S) 
    \nonumber\\
    & \quad +120e^{2\pi i z_2} (i +\pi z_2)(11z_1 +23z_2 +15z_3 +9S)
    \biggl].
\label{eq:Piqu}
\end{align}
Here $\Pi_{i}$ just denote the perturbative expansions of period integrals described 
in (\ref{Picl}) and we have added leading corrections due to the one-instanton 
${\cal O}(e^{-2\pi {\rm Im}(z)})$ effect from the A-model perspective, which is 
sufficient to exemplify the uplifting mechanism realizing F-theory vacua with small 
superpotential.

Accordingly, the K$\ddot{{\textrm{a}}}$hler potential (\ref{expliK}) is also deformed 
into the following expression:
\begin{align}
\widetilde{K}=K
&-\frac{{\rm Im}(z_1)\left(1+2\pi{\rm Im}(z_1)\right)}{2\pi^3}e^{-2\pi {\rm Im}
(z_1)}\cos (2\pi {\rm Re}(z_1)) \nonumber\\
&+\frac{120\left(1+2\pi{\rm Im}(z_2)\right)\left(11{\rm Im}(z_1)+23{\rm Im}(z_2)
+15{\rm Im}(z_3)+9{\rm Im}(S)\right)}{2\pi^3}e^{-2\pi {\rm Im}(z_2)}
\cos (2\pi {\rm Re}(z_2)) \nonumber\\
&+\frac{\left(1+2\pi{\rm Im}(z_3)\right)\left(2{\rm Im}(z_1)+{\rm Im}(z_3)\right)}
{2\pi^3}e^{-2\pi {\rm Im}(z_3)}\cos (2\pi {\rm Re}(z_3)) \nonumber\\
&+\frac{2\left(1+2\pi{\rm Im}(S)\right)\left({\rm Im}(z_1)+{\rm Im}(z_3)\right)}
{2\pi^3}e^{-2\pi {\rm Im}(S)}\cos (2\pi {\rm Re}(S)).
\end{align}

As mentioned before, we rely on a numerical calculation to find out desirable F-theory 
flux vacua with exponentially small superpotential. More precisely, in order to demonstrate
the non-perturbative uplifting mechanism, we utilized the Mathematica to solve the 
intricate differential system (\ref{vsol}), and analyzed the vacuum structure of the 
present model under the following range of background fluxes: 
\begin{align}
    -20 \leq n_{6},n_7,n_8,n_9,n_{10},n_{11}\leq 20.
\end{align}
The tadpole cancellation 
condition (\ref{tad}) whose explicit form is given by
\begin{align}
\begin{split}
702 &= n_{D3}+n_1n_{16}+n_2n_{12}+n_3n_{13}+n_4n_{14}+n_5n_{15}+\frac{1}{7}n_6n_{10}
+\frac{8}{7}n_6n_{11}+n_6^2-\frac{1}{2}n_7n_{10} \\
& \ \ \ +\frac{1}{2}n_7n_{11}+\frac{1}{4}n_7^2+\frac{5}{14}n_8n_{10}
-\frac{9}{14}n_8n_{11}-n_6n_8-\frac{3}{4}n_7n_8+\frac{1}{4}n_8^2-\frac{5}{7}n_6n_9
+\frac{3}{14}n_8n_9
\end{split}
\end{align}
must be also satisfied. Note that we have also restricted ourselves to a small 
 number of mobile D3-branes as $0 \leq n_{D3} \leq 10$, in order to legitimately 
exclude the effect of backreaction to the space-time.

As a result, we found that there are 18 perturbatively flat solutions of the type given by 
(\ref{Floc}) in the present setup, and after including the instanton corrections to the 
potential, 3 out of them remain to satisfy the differential system (\ref{vsol}) and realize 
desirable class of flux vacua with small superpotential.\footnote{To guarantee the convergence 
of the expansion around a large complex structure point of the period integrals and the 
superpotential, we picked up numerical solutions satisfying the conditions ${\textrm{Im}} (z_i) >1$ only.}
Here we denote these 3 different solutions as Vacua A, B, and C, whose background data such 
as the $G_4$-flux quanta and number of mobile D3-branes have been determined as follows:

\begin{table}[h]
    \centering
    \begin{tabular}{|c|c|c|} \hline
Vacuum & Set of fluxes $(n_6, n_7, n_8, n_9, n_{10}, n_{11})$ & $n_{D3}$\\ \hline \hline
A & $(-10, -8, 12, 7, 15, -8)$ & 2 \\ \hline
B &  $(-9, -8, 14, 0, 11, -11)$ & 5\\ \hline
C & $(-15, 8, 6, 20, -4, -8)$ & 6\\  \hline
    \end{tabular}
\label{tab:flux}
\end{table}

\noindent 
The explicit values of the superpotential $W_0 \equiv \langle e^{K/2}W\rangle$, 
stabilized moduli fields and their mass squareds at  each of the flux vacua are 
summarized in Tables~\ref{tab:vev} and~\ref{tab:mass}, where $V$ and ${\cal V}$
are the scalar potential and overall volume of the background respectively.
\begin{table}
    \centering
    \begin{tabular}{|c|c|c|c|c|c|} \hline
Vacuum &  $z_1$  & $z_2$ & $z_3$ & $S$ & $|W_0|$ \\ \hline \hline
A & $1.95 i$ & $2.60 i$ & $5.86 i$ & $4.56 i$ & $6.75\times 10^{-9}$ \\ \hline
B & $1.35 i$ & $1.97 i$ & $4.06 i$ & $4.06 i$ & $6.11\times 10^{-7}$ \\ \hline
C & $2.41 i$ & $1.81 i$ & $1.20 i$ & $2.71 i$ & $2.50\times 10^{-6}$ \\ \hline
    \end{tabular}
    \caption{Explicit values of the stabilized moduli fields and small flux superpotential 
    in $M_{\rm Pl}=1$ unit.}
    \label{tab:vev}
\end{table}
\begin{table}
    \centering
    \begin{tabular}{|c|c|} \hline
Vacuum    & Eigenvalues of mass matrix $\partial_I \partial_J V \times {\cal V}^{2}$
   \\ \hline \hline
A & (24.7, 24.7, 4.86, 4.86, 0.634, 0.634, $9.79\times 10^{-14}$, $9.65\times 10^{-14}$) \\ \hline
B & (42.5, 42.5, 8.76, 8.76, 1.33, 1.33, $4.68\times 10^{-10}$, $4.56\times 10^{-10}$)\\ \hline
C & (61.9, 61.9, 15.2, 15.2, 0.765, 0.765, $1.30\times 10^{-8}$, $1.27\times 10^{-8}$)\\ \hline
    \end{tabular}
    \caption{Mass squareds of canonically normalized moduli fields in $M_{\rm Pl}=1$ unit.}
    \label{tab:mass}
\end{table}
In the case of Vacuum A, the perturbatively flat direction along (\ref{eq:flatex}) has 
been stabilized due to a small non-perturbative contribution to the superpotential of 
a racetrack-type given by
\begin{align}
    W = \frac{1}{4\pi^2}\biggl[ -44e^{\frac{6\pi i S}{7}} +2040e^{\frac{8\pi i S}{7}} 
    +49e^{2\pi i S} -68 e^{\frac{18\pi i S}{7}}\biggl],
\label{eq:Weff}
\end{align}
and the perturbative shift symmetry is broken down to a discrete subgroup. Here we 
used the fact that orthogonal directions to the flat direction are all heavy and 
can be integrated out. Note that light modes originating from the uplifted flat direction
can be involved with the dynamics of K$\ddot{{\textrm{a}}}$hler moduli fields, as discussed 
in\cite{Demirtas:2019sip,Demirtas:2020ffz,Blumenhagen:2020ire}. It is worth noting that 
one can straightforwardly generalize our analysis to include higher-order instanton 
corrections as well. Especially, we confirmed that the above numerical solutions with tiny 
flux superpotential continue to be stable against the next instanton corrections.

\section{Conclusions and Discussions}
\label{sec:con}

Toward an explicit realization of our four-dimensional physics within the framework of the
general concept of F-theory, we explored the possibility of the realization of F-theory 
flux compactifications with exponentially small superpotential. It has been recently pointed
out in \cite{Demirtas:2019sip,Demirtas:2020ffz} that non-perturbative corrections to complex 
structure moduli fields can naturally give rise to tiny flux superpotential in Type IIB 
string theory on Calabi-Yau threefolds. Since the smallness of the flux superpotential plays 
a crucial role in KKLT-type construction for de Sitter space, we examined whether such 
a simple but broadly applicable method can be also realized in F-theory compactified on
Calabi-Yau fourfolds.  

Generalizing the Type IIB setups analyzed in \cite{Demirtas:2019sip,Demirtas:2020ffz}
into the frameworks of F-theory compactifications, we clarified that a restricted choice 
of $G_4$-flux components reduces the flux superpotential into a quite simple 
form of homogeneous of degree two, and a class of supersymmetric F-theory vacua can 
perturbatively exist along with one flat direction. Then we determined one-instanton 
corrections to the potential of a particular example by utilizing the techniques of 
mirror symmetry, and numerically investigated its vacuum structure at the non-perturbative 
level.

As a result, we numerically confirmed that flat directions of perturbative vacua of the 
model can be lifted appropriately and remaining modulus acquired a small mass, along with 
a desired tiny superpotential. Although our explicit demonstration of this uplifting 
mechanism has the potential for tremendous impact on the implementation of KKLT 
construction in a broad range of F-theory frameworks, rigorous calculation about K\"ahler 
moduli stabilization in F-theory requires more precise understanding about strong dynamics 
of seven-branes, remaining an open problem.

From a statistical point of view, it would be interesting to clarify to what extent 
flux vacua with small superpotential distribute inside the string/F-theory landscape. 
One possible approach to address this intriguing subject is to promote discrete $G_4$ 
fluxes to continuous parameters and attempt to obtain a reasonable estimate for the 
numbers of possible F-theory vacua, as initiated 
in \cite{Ashok:2003gk,Douglas:2003um,Denef:2004ze} for Type IIB flux compactifications. 

Moreover, the authors of \cite{Demirtas:2020ffz,Blumenhagen:2020ire} recently applied 
the non-perturbative uplifting mechanism to the conifold region of 
Calabi-Yau moduli space and explicitly found conifold vacua with small flux 
superpotential. Although a comprehensive study about global structure of moduli space 
of Calabi-Yau fourfolds has not yet been fully elucidated, it is also fascinating to 
extend our F-theory setup into other corners of the landscape in the future.

\subsection*{Acknowledgements}

We would like to thank K. Ohta for useful discussions and comments, and especially to the
anonymous referee for a careful reading of the manuscript and constructive suggestions. 
H. O. was supported in part by JSPS KAKENHI Grant Numbers JP19J00664 and JP20K14477.

\appendix

\section{Another type of perturbative solution}
\label{subsec:Noninv}

Here we comment on the existence of another type of perturbatively flat vacua purposefully 
omitted from the main text. Although the matrix $C_{ab}$ has been assumed to be invertible 
throughout this paper, generically it is also possible to realize desirable solutions 
equipped with flat directions, even when $\textrm{Det} (C_{ab})=0$. In the case of the 
example described in Section \ref{subsec:P11169set}, one can easily show that there exist 
two more perturbative solutions depending on the choice of $G_4$-fluxes as follows:

\begin{itemize}
    \item 
    Under the condition
    \begin{align}
    n_{7} &= 0,
    \nonumber\\
    n_8 &= -3n_{10} +4n_{11}+n_6+\frac{-5n_{11}^2+\sqrt{((6n_{10}-5n_{11})^2-4n_{10}n_6)((2n_{10}-n_{11})^2-4n_{10}n_6)}}{4n_{10}},
    \nonumber\\
    n_9 &= 3n_{10} -\frac{n_{11}}{2},
\end{align}
there exists a solution to (\ref{vsol}) satisfying $\textrm{Det} (C_{ab})= \widetilde{C}_a =0$ 
with an additional flat direction given by 
\begin{align}
    \left(
    \begin{array}{c}
         z_2  \\
         z_3   
    \end{array}
    \right)
    =z_1
    \left(
    \begin{array}{c}
        \frac{-12 n_{10}^2 +16n_{10}n_{11}- 5n_{11}^2 - 4 n_{10} n_6 + 
   \sqrt{((6 n_{10} - 5 n_{11})^2 - 4 n_{10} n_6) ((-2 n_{10} + n_{11})^2 - 
      4 n_{10} n_6)}}{4n_{10} (4 n_{10} - 3 n_{11})}  \\
         \frac{(-2 n_{10} + n_{11})^2 - 4 n_{10} n_6 + 
   \sqrt{((6 n_{10} - 5 n_{11})^2 - 4 n_{10} n_6) ((-2 n_{10} + n_{11})^2 - 
      4 n_{10} n_6)}}{4 (4 n_{10} - 3 n_{11}) (n_{10} - n_{11})}
    \end{array}
    \right),
\end{align}
as well as the obvious flat direction parametrized by $S$.

    \item 
    Under the condition
    \begin{align}
    n_{6} &= -\frac{n_7 \left(16 n_{10}^2 + n_{10}
    (-44 n_{11}+ 5 n_{7}) + 3 n_{11} (8 n_{11} + 5 n_{7})\right)}{
 4 (4 n_{10} - 3 n_{11})^2},
    \nonumber\\
    n_8 &= 2n_6,
    \nonumber\\
    n_9 &= \frac{n_{10}+8n_{11}}{5},
\end{align}
there exists a solution to (\ref{vsol}) satisfying $\textrm{Det} (C_{ab})=0$ 
with a flat direction given by 
\begin{align}
    \left(
    \begin{array}{c}
         z_2  \\
         S   
    \end{array}
    \right)
    =(z_1+z_3)
    \left(
    \begin{array}{c}
         -\frac{5n_7}{8n_{10}-6n_{11}}
         \\
           \frac{8 n_{10}^2 + n_{10}
    (-22 n_{11}+ 5 n_{7}) + 3 n_{11} (4 n_{11} + 5 n_{7})}{2(4 n_{10} - 3 n_{11})^2}  
    \end{array}
    \right),
\end{align}
as well as a direction given by $z_1-z_3$ locus.
\end{itemize}
The non-perturbative uplifting mechanism explicitly demonstrated in 
Section \ref{subsec:P11669qu} is presumed to be straightforwardly applied 
to this type of solutions.



\begin{thebibliography}{99}
\parskip=-2pt


\bibitem{Giddings:2001yu}
S.~B.~Giddings, S.~Kachru and J.~Polchinski,
Phys. Rev. D \textbf{66} (2002), 106006
[arXiv:hep-th/0105097 [hep-th]].

\bibitem{Kachru:2003aw}
S.~Kachru, R.~Kallosh, A.~D.~Linde and S.~P.~Trivedi,
Phys. Rev. D \textbf{68} (2003), 046005
[arXiv:hep-th/0301240 [hep-th]].

\bibitem{Balasubramanian:2005zx}
V.~Balasubramanian, P.~Berglund, J.~P.~Conlon and F.~Quevedo,
JHEP \textbf{03} (2005), 007
[arXiv:hep-th/0502058 [hep-th]].

\bibitem{Demirtas:2019sip}
M.~Demirtas, M.~Kim, L.~Mcallister and J.~Moritz,
``Vacua with Small Flux Superpotential,''
Phys. Rev. Lett. \textbf{124} (2020) no.21, 211603
[arXiv:1912.10047 [hep-th]].

\bibitem{Demirtas:2020ffz}
M.~Demirtas, M.~Kim, L.~Mcallister and J.~Moritz,
``Conifold Vacua with Small Flux Superpotential,''
[arXiv:2009.03312 [hep-th]].

\bibitem{Blumenhagen:2020ire}
R.~\'Alvarez-Garc\'\i{}a, R.~Blumenhagen, M.~Brinkmann and L.~Schlechter,
``Small Flux Superpotentials for Type IIB Flux Vacua Close to a Conifold,''
[arXiv:2009.03325 [hep-th]].

\bibitem{Vafa:1996xn}
C.~Vafa,
``Evidence for F theory,''
Nucl. Phys. B \textbf{469}, 403-418 (1996)
[arXiv:hep-th/9602022 [hep-th]].

\bibitem{Denef:2008wq}
F.~Denef,
``Les Houches Lectures on Constructing String Vacua,''
Les Houches \textbf{87}, 483-610 (2008)
[arXiv:0803.1194 [hep-th]].

\bibitem{Gukov:1999ya}
S.~Gukov, C.~Vafa and E.~Witten,
Nucl. Phys. B \textbf{584}, 69-108 (2000)
[erratum: Nucl. Phys. B \textbf{608}, 477-478 (2001)]
[arXiv:hep-th/9906070 [hep-th]].

\bibitem{Becker:1996gj}
K.~Becker and M.~Becker,
``M theory on eight manifolds,''
Nucl. Phys. B \textbf{477}, 155-167 (1996)
[arXiv:hep-th/9605053 [hep-th]].

\bibitem{Sethi:1996es}
S.~Sethi, C.~Vafa and E.~Witten,
``Constraints on low dimensional string compactifications,''
Nucl. Phys. B \textbf{480}, 213-224 (1996)
[arXiv:hep-th/9606122 [hep-th]].

\bibitem{Dasgupta:1999ss}
K.~Dasgupta, G.~Rajesh and S.~Sethi,
``M theory, orientifolds and G - flux,''
JHEP \textbf{08}, 023 (1999)
[arXiv:hep-th/9908088 [hep-th]].

\bibitem{Haack:2001jz}
M.~Haack and J.~Louis,
``M theory compactified on Calabi-Yau fourfolds with background flux,''
Phys. Lett. B \textbf{507}, 296-304 (2001)
[arXiv:hep-th/0103068 [hep-th]].

\bibitem{Alim:2009bx}
M.~Alim, M.~Hecht, H.~Jockers, P.~Mayr, A.~Mertens and M.~Soroush,
``Hints for Off-Shell Mirror Symmetry in type II/F-theory Compactifications,''
Nucl. Phys. B \textbf{841}, 303-338 (2010)
[arXiv:0909.1842 [hep-th]].

\bibitem{Grimm:2009ef}
T.~W.~Grimm, T.~W.~Ha, A.~Klemm and D.~Klevers,
``Computing Brane and Flux Superpotentials in F-theory Compactifications,''
JHEP \textbf{04}, 015 (2010)
[arXiv:0909.2025 [hep-th]].

\bibitem{Jockers:2009ti}
H.~Jockers, P.~Mayr and J.~Walcher,
``On N=1 4d Effective Couplings for F-theory and Heterotic Vacua,''
Adv. Theor. Math. Phys. \textbf{14}, no.5, 1433-1514 (2010)
[arXiv:0912.3265 [hep-th]].

\bibitem{Grimm:2010gk}
T.~W.~Grimm, A.~Klemm and D.~Klevers,
``Five-Brane Superpotentials, Blow-Up Geometries and SU(3) Structure Manifolds,''
JHEP \textbf{05}, 113 (2011)
[arXiv:1011.6375 [hep-th]].

\bibitem{Honma:2015iza}
Y.~Honma and M.~Manabe,
``Open Mirror Symmetry for Higher Dimensional Calabi-Yau Hypersurfaces,''
JHEP \textbf{03}, 160 (2016)
[arXiv:1507.08342 [hep-th]].

\bibitem{Honma:2013hma}
Y.~Honma and M.~Manabe,
``Exact Kahler Potential for Calabi-Yau Fourfolds,''
JHEP \textbf{05}, 102 (2013)
[arXiv:1302.3760 [hep-th]].

\bibitem{Honma:2017uzn}
Y.~Honma and H.~Otsuka,
``On the Flux Vacua in F-theory Compactifications,''
Phys. Lett. B \textbf{774}, 225-228 (2017)
[arXiv:1706.09417 [hep-th]].

\bibitem{Honma:2019gzp}
Y.~Honma and H.~Otsuka,
``F-theory Flux Vacua and Attractor Equations,''
JHEP \textbf{04}, 001 (2020)
[arXiv:1910.10725 [hep-th]].

\bibitem{Batyrev:1993dm} 
  V.~V.~Batyrev,
  ``Dual Polyhedra and Mirror Symmetry for Calabi-Yau Hypersurfaces in Toric Varieties,''
  alg-geom/9310003.

\bibitem{Batyrev:1994pg} 
  V.~V.~Batyrev and L.~A.~Borisov,
  ``On Calabi-Yau complete intersections in toric varieties,''
  alg-geom/9412017.

\bibitem{Hosono:1993qy}
S.~Hosono, A.~Klemm, S.~Theisen and S.~T.~Yau,
``Mirror symmetry, mirror map and applications to Calabi-Yau hypersurfaces,''
Commun. Math. Phys. \textbf{167}, 301-350 (1995)
[arXiv:hep-th/9308122 [hep-th]].

\bibitem{Hosono:1994ax}
S.~Hosono, A.~Klemm, S.~Theisen and S.~T.~Yau,
``Mirror symmetry, mirror map and applications to complete intersection Calabi-Yau spaces,''
AMS/IP Stud. Adv. Math. \textbf{1}, 545-606 (1996)
[arXiv:hep-th/9406055 [hep-th]].

\bibitem{Ashok:2003gk}
S.~Ashok and M.~R.~Douglas,
JHEP \textbf{01} (2004), 060
[arXiv:hep-th/0307049 [hep-th]].

\bibitem{Douglas:2003um}
M.~R.~Douglas,
JHEP \textbf{05} (2003), 046
[arXiv:hep-th/0303194 [hep-th]].

\bibitem{Denef:2004ze}
F.~Denef and M.~R.~Douglas,
JHEP \textbf{05} (2004), 072
[arXiv:hep-th/0404116 [hep-th]].


\end{thebibliography}
\end{document}